\documentclass{article}
\usepackage{amsmath}
\usepackage{amsthm}
\newtheorem{theorem}{Theorem}
\usepackage{verbatim}
\usepackage{amssymb}
\usepackage{algorithm}
\usepackage{algorithmic}
\newtheorem{definition}{Definition}
\usepackage{caption}
\usepackage{graphicx} 
\makeatletter
\newenvironment{breakablealgorithm}
  {% \begin{breakablealgorithm}
   \begin{center}
     \refstepcounter{algorithm}%
     \hrule height.8pt depth0pt \kern2pt% Top rule
     \renewcommand{\caption}[2][default]{\vskip2pt\small\textbf{\ALG@name~\thealgorithm:} ##2\vskip5pt\hrule\vskip2pt}
  }
  {% \end{breakablealgorithm}
     \kern2pt\hrule\relax% Bottom rule
   \end{center}
  }
\makeatother
\usepackage[preprint]{neurips_2026}

\usepackage[utf8]{inputenc} % allow utf-8 input
\usepackage[T1]{fontenc}    % use 8-bit T1 fonts
\usepackage{hyperref}       % hyperlinks
\usepackage{url}            % simple URL typesetting
\usepackage{booktabs}       % professional-quality tables
\usepackage{amsfonts}       % blackboard math symbols
\usepackage{nicefrac}       % compact symbols for 1/2, etc.
\usepackage{microtype}      % microtypography
\usepackage{xcolor}         % colors

\title{Learning Market Competition in Shared Spectrum: A Multi-Agent Reinforcement Learning Approach}

\author{%
  Qixuan Zai, Randall Berry\\
  Department of Electrical and Computer Engineering\\
  Northwestern University, Evanston, IL 60208, USA\\
  \texttt{Emails:\{QixuanZai2028@u.,rberry@\}northwestern.edu} \\
}

\begin{document}

\maketitle

\begin{abstract}
This paper investigates market competition among wireless service providers (SPs) that serve customers using shared spectrum. Prior work has analyzed such markets through models of competition with congestible resources, capturing both the congestion-sensitive nature of wireless spectrum and the effects of spectrum sharing on service quality. These models typically assume that the market demand function is known, enabling SPs to optimize pricing or quantity decisions under either Bertrand or Cournot competition. In contrast, we consider a setting in which the demand function is initially unknown and must be learned over time. We model this learning process using multi-agent reinforcement learning (MARL), allowing competing SPs to learn market dynamics while adapting their competitive strategies. Although MARL has shown strong performance in a variety of economic settings, recent work has demonstrated that it can also give rise to tacit collusion among self-interested agents. We therefore examine whether similar collusive behavior emerges in shared-spectrum markets and how its prevalence depends on the mode of competition (price versus quantity) and the choice of MARL algorithm. Our results provide insight into the interaction between learning dynamics, market structure, and spectrum sharing, with implications for both wireless market design and the deployment of learning-enabled decision-making systems.
\end{abstract}

\section{Introduction}
The rapid growth of wireless services has intensified demand for radio spectrum, motivating spectrum-sharing frameworks that allow multiple service providers (SPs) to operate over common frequency bands. Examples include expanded unlicensed access in the 6~GHz band and the Citizens Broadband Radio Service (CBRS) framework in the 3.5~GHz band. Compared with exclusive licensing, spectrum sharing can improve spectrum utilization and accommodate growing traffic demands. However, because multiple providers compete for access to a common spectrum resource, the service quality experienced by each provider depends on the spectrum usage of others, creating strategic interactions that are naturally modeled as non-cooperative congestion games~\cite{rosenthal1973congestion,johari2004efficiency}. A substantial body of work has analyzed competition in shared-spectrum markets using economic models with congestible resources. In these models, customers choose providers based on a delivered price consisting of the announced service price and a congestion cost determined by spectrum utilization. This framework captures the coupling between wireless interference and market competition and has been applied to a wide range of spectrum-sharing scenarios, including intermittent spectrum availability, licensed shared access, and prioritized spectrum sharing~\cite{Q2,Q3,Q1}. These models have provided important insights into equilibrium prices, market shares, profits, consumer surplus, and social welfare under both Bertrand (price) and Cournot (quantity) competition~\cite{bertrand1883theorie,cournot1838recherches}.

Despite their broad applicability, existing congestion-based market models generally assume that market demand is known to all competing service providers, enabling analytical characterization of equilibrium strategies~\cite{johari2004efficiency}. In practice, however, providers rarely possess complete knowledge of demand, which evolves over time as user preferences, competing technologies, and market conditions change. Instead, providers must repeatedly interact with the market, infer demand from customer responses, and adapt their competitive strategies over time. Reinforcement learning (RL) provides a natural framework for sequential decision-making under such uncertainty~\cite{sutton2018reinforcement}. Because multiple providers learn simultaneously in a shared competitive environment, the problem is more appropriately modeled using multi-agent reinforcement learning (MARL), which enables self-interested agents to learn effective strategies directly from repeated interactions without requiring prior knowledge of demand models or competitors' actions. However, while MARL provides a natural framework for learning unknown demand, recent studies have also shown that learning agents may develop tacitly collusive pricing strategies without explicit communication~\cite{calvano2020algorithmic,klein2021autonomous}, raising important concerns for competition policy and AI-enabled markets.

To date, however, the MARL collusion literature has focused almost exclusively on conventional oligopoly models in which firms compete through prices or quantities without congestion effects \cite{calvano2020algorithmic,klein2021autonomous,fish2024algorithmic}. Conversely, the wireless spectrum-sharing literature has largely assumed rational firms with complete knowledge of market demand and has analyzed market equilibria under this assumption \cite{johari2004efficiency,Q2}. As a result, little is known about how learning dynamics interact with congestion externalities in shared-spectrum markets\cite{Zai2026MultiAgentRL}. Congestion fundamentally alters firms' incentives because each provider's actions affect not only market prices but also the quality of service experienced by all providers' customers, creating strategic interactions through a shared resource \cite{rosenthal1973congestion,monderer1996potential,johari2004efficiency}.  Whether these additional congestion-induced strategic interactions amplify or suppress the emergence of tacit collusion, and how they affect market efficiency and consumer welfare, remains largely unexplored.

To the best of our knowledge, this is the first work to integrate congestion-based wireless spectrum markets with multi-agent reinforcement learning, enabling service providers to learn unknown demand while quantifying the emergence of tacit collusion under both Bertrand and Cournot competition. Building on the congestion-market framework, we replace the assumption of known demand with adaptive learning. We formulate repeated Bertrand and Cournot competition over shared spectrum as MARL problems in which competing service providers adapt their pricing or quantity strategies in the face of unknown demand. While a linear inverse demand function is assumed to derive analytical benchmarks and characterize equilibrium behavior, the learning agents are not given this demand model. Instead, they must learn effective pricing or quantity strategies solely through repeated interactions with the environment, without prior knowledge of the underlying demand model. We investigate whether competing providers converge to competitive or collusive outcomes, how these outcomes depend on the mode of competition (price versus quantity), and how different MARL algorithms influence the resulting market behavior.
The main contributions of this paper are summarized as follows.
\begin{itemize}
\item We develop a multi-agent reinforcement learning framework for congestion-based wireless markets that replaces the classical assumption of known demand with adaptive learning.
\item We formulate both Bertrand and Cournot competition over shared spectrum as a Markov game, allowing competing service providers to simultaneously learn demand and optimize their competitive strategies.
\item We compare multiple MARL algorithms and characterize how learning algorithms affect convergence, market efficiency, and the emergence of tacit collusion. 
\item We demonstrate that congestion fundamentally changes the emergence of tacit collusion relative to conventional oligopoly models, providing new insights into the interaction between learning dynamics and wireless market structure. 
\end{itemize}

Our results show that congestion fundamentally changes the learning dynamics and strategic behavior observed in conventional MARL pricing models. The interaction between shared-spectrum congestion, market competition, and adaptive learning can either strengthen or weaken collusive behavior depending on the competition model and learning algorithm. These findings contribute to both the economics of wireless spectrum sharing and the growing literature on learning-enabled strategic decision making, with implications for the design of future autonomous wireless markets.

The remainder of this paper is organized as follows. Section 2 establishes the formal system model for both the Cournot and Bertrand congestion frameworks. Section 3 details the multi-agent deep reinforcement learning algorithms deployed to solve these equilibrium under dynamic conditions. Section 4 presents comprehensive experimental results validating the varying degrees of collusion observed across both game structures. Finally, Section 5 concludes the paper and discusses future research directions.

\section{System Model}

We consider a shared-spectrum market consisting of one incumbent service provider and $N$ entrant providers. Our model extends the congestion-market framework of \cite{berry2020value,Q3} by replacing the assumption of complete demand information with repeated learning. While the market follows the same congestion and Wardrop equilibrium structure as prior work, competing providers do not know the underlying demand function and must instead learn effective strategies from observed market outcomes. We assume that the incumbent has its own licensed band of spectrum with bandwidth $B_e$ and entrants share access to a common band of bandwidth $W_e$.\footnote{This models a setting in which the shared band enables new entrants to compete against an existing incumbent.  We focus on one incumbent for simplicity of the presentation, but this type of model can be readily extended to settings with multiple incumbents as in \cite{Q2}.}

The incumbent is denoted as SP 1, while the entrants are indexed from 2 to $N+1$. All SPs compete for a common pool of non-atomic customers. Let $x_{i}$ denote the amount of users served by SP $i\;(1\le i\le N+1)$. The quality of service experienced by users is determined by congestion on the spectrum band, modeled through a congestion cost $g(X/Y)$, where $X$ denotes the total amount of users on the band, $Y$ is the bandwidth, and $g(\cdot)$ is a convex increasing function, satisfying $g(0)=0$. We further assume that $g(\cdot)$ is linear, which makes the problem more tractable. The average congestion cost of the incumbent can be expressed as 
\begin{equation}
    \begin{split}
    \hat{g}_{\mathrm{in}}\left(x_{1}\right) \triangleq g\left(\frac{x_{1}}{B_e}\right).
    \label{eq:congest_in}
    \end{split}
\end{equation}
The average congestion cost for the entrant service is 
\begin{equation}
    \begin{split}
    \hat{g}_{\mathrm{en}}\left(\mathbf{x}\right) \triangleq g\left(\frac{x_{-1}}{W_e}\right),
    \label{eq:congest_en}
    \end{split}
\end{equation}
where $x_{-1} = x_{2}+\cdots+x_{N+1}$ is the total user amount of entrants. 

Each SP collects a fixed service price $p_i$ from each user it serves, so its revenue is $x_i p_i$. The delivered price of a SP $i$ is the sum of its service price and the congestion of its service, i.e., $p_1 + \hat{g}_\mathrm{in}(x_{1})$ for $i=1$ or $p_i + \hat g_\mathrm{en}(\mathbf x)$ for $2\le i\le N+1$. 
Customer demand is characterized by an unknown inverse demand function $P(X)$, where $P(X)$ denotes the maximum delivered price that can be sustained when the total demand is $X$.
 Note that the demand function depends on the delivered price, which models customers that care about both the price they pay for service and the quality of service (represented by the congestion cost). The amount of users served by each SP and the announced prices satisfy the Wardrop equilibrium conditions, that is, the delivered prices of all SPs that are serving users are the same and are the maximum price that can be accepted by all the customers serviced \cite{wardrop1952some}. Assuming all SP's are serving customers, this can be expressed as 
\begin{equation}
\begin{aligned}
P(X) &= p_1 + \hat{g}_{\mathrm{in}}(x_1), \quad \text{and} \\
P(X) &= p_i + \hat{g}_{\mathrm{en}}(\mathbf{x}), \quad 2 \le i \le N+1,
\end{aligned}
\label{eq:wardropeq}
\end{equation}
where $X=x_1+x_2+\cdots+x_{N+1}$ is the total demand. If a SP $i$ is not serving any customers ($x_i=0$), then its delivered price must be greater than or equal to $P(X)$. 

The market is characterized by an unknown inverse demand function $P(X)$, which is not revealed to the learning agents. For analytical benchmarking, however, we specialize to a linear inverse demand function when deriving the theoretical equilibria presented below. In each interaction, providers observe the resulting market outcome (i.e., their revenue) but do not observe the underlying demand function or competitors' private decision process.

Within this common market framework, we consider two classical competition models. Under Cournot competition, providers compete by selecting service quantities, whereas under Bertrand competition they compete by selecting prices. These models provide complementary settings for evaluating how learning dynamics interact with congestion.

\subsection{Cournot Congestion Competition}

Under Cournot competition, each service provider chooses the quantity of users it wishes to serve, while the corresponding market prices are determined endogenously through the Wardrop equilibrium conditions in (\ref{eq:wardropeq}). This models providers that compete by controlling market share rather than directly setting prices. The resulting prices depend on the aggregate quantities selected by all providers together with the congestion experienced on the licensed and shared spectrum bands.

In the classical complete-information setting, providers are assumed to know the market demand function and compute a Nash equilibrium analytically. Under the assumptions considered here, the resulting congestion game belongs to the class of potential games, which guarantees the existence of a pure-strategy Nash equilibrium and provides favorable convergence properties under best-response dynamics~\cite{monderer1996potential}. In contrast, this paper considers an incomplete-information setting in which providers do not know the underlying demand function. Instead, they repeatedly interact with the market, observe the resulting outcomes, and learn effective quantity strategies through multi-agent reinforcement learning.

For a given strategy profile of the competing providers, service provider $i$ solves
\begin{equation}
\max_{x_i \ge 0} \; x_i p_i,
\label{eq:cournot_opt}
\end{equation}
where $p_i$ is determined implicitly through the Wardrop equilibrium conditions in (\ref{eq:wardropeq}). Throughout the paper, this complete-information equilibrium serves as the theoretical benchmark against which the learned MARL policies are compared.

For analytical benchmarking, we specialize to a linear inverse demand function,
\[
P(X)=1-X,
\]
together with the linear congestion model introduced above.

\subsection{Bertrand Competition}

Under Bertrand competition, service providers compete by selecting service prices rather than customer quantities. The resulting customer allocation is determined endogenously through the Wardrop equilibrium conditions in (\ref{eq:wardropeq}), which couple the announced prices, congestion costs, and total market demand. This model captures markets in which providers compete directly on price while customers choose the provider offering the lowest delivered price.

Unlike the Cournot congestion game, Bertrand competition exhibits substantially different strategic behavior. Classical Bertrand theory predicts that firms offering homogeneous products compete prices down to marginal cost, yielding the well-known Bertrand paradox~\cite{bertrand1883theorie,tirole1988theory}. In contrast, recent work has demonstrated that reinforcement-learning agents may instead converge to tacitly collusive pricing strategies without explicit communication~\cite{calvano2020algorithmic,klein2021autonomous}. A primary objective of this paper is to determine how congestion and spectrum sharing influence this learned pricing behavior.

Each provider selects its announced service price to maximize its revenue, while the resulting customer allocation satisfies the Wardrop equilibrium conditions. Specifically, service provider $i$ solves
\begin{equation}
\max_{p_i \ge 0} \; x_i p_i,
\label{eq:bertrand_opt}
\end{equation}
where the demand allocated to each provider depends on the prices announced by all competing providers together with the resulting congestion costs. As in the Cournot model, the complete-information equilibrium serves as the analytical benchmark against which the learned MARL policies are evaluated.

For analytical benchmarking, we assume the same linear inverse demand function introduced previously,
\[
P(X)=1-X,
\]
while maintaining the linear congestion model. 

The theoretical Bertrand equilibrium provides the competitive benchmark for the empirical learning results presented in Section~IV. Deviations from this benchmark, particularly persistent prices above the competitive equilibrium, indicate the emergence of tacit collusive behavior among the learning agents. Comparing these outcomes with those obtained under Cournot competition allows us to isolate the influence of congestion on algorithmic pricing and market efficiency.

\subsection{Analytical Benchmarks}

The Nash equilibria of congestion-based Cournot and Bertrand markets under complete information have been extensively studied in the literature~\cite{johari2004efficiency,Q3}, we summarize these in Table~II assuming a linear inverse demand, $P(X)= 1-X$ and linear congestion costs for the case where all SPs are serving customers.\footnote{The expressions when some SPs are not serving customer can also be derived, but are omitted to simplify the presentation.}

As we have noted MARL frameworks have been observed to result in collusive behavior. As a benchmark for this, we also determine the collusive outcome in which all SPs coordinate on prices or quantities to maximize the total revenue (summed across firms).  This outcome is also shown in Table I.\footnote{This theoretical benchmark assumes a continuous action space; slight deviations may occur in discrete settings. Additionally, we assume that the number of entrants $2\leq N$ and index $1<i\leq N+1$.} Unlike the Nash equilibrium, the collusive benchmark is not intended as a prediction of rational strategic behavior. Rather, it represents the market outcome that would arise under perfect coordination. Although such explicit coordination is not permitted, this benchmark provides a useful upper bound for evaluating the extent of tacit collusion learned by MARL agents.

\begin{table*}[t]
\centering
\caption{Summary of Analytical Results Across Different Market Models}
\label{t1112}
\setlength{\tabcolsep}{0.005\linewidth} % Optimal spacing for a two-column spanning layout
\begin{tabular}{|c| c| c| c|}
\toprule
{\tt Metric} & {\tt Collusive Model} & {\tt Bertrand Nash} & {\tt Cournot Nash } \\
\midrule
$x_1$ & $\frac{B_{e}}{2(B_e+W_e+1)}$ & $\frac{B_e}{2(B_e+W_e+1)}$ & $\frac{B_{e} \left( \left(N + 1\right) \left(W_{e} + 1\right) - N W_{e} \right)}{2\left(B_{e} + 1\right) \left(N + 1\right) \left(W_{e} + 1\right)- B_{e} N W_{e}}$ \\
\rule{0pt}{6ex} $x_i$ & $\frac{W_{e}}{2N(B_e+W_e+1)}$ & $\frac{W_e(B_e+2W_e+2)}{2N(W_e+1)(B_e+W_e+1)}$ & $\frac{W_{e} \left(B_{e} + 2\right)}{2\left(B_{e} + 1\right) \left(N + 1\right) \left(W_{e} + 1\right)- B_{e} N W_{e}}$ \\
\rule{0pt}{6ex} $p_1$ & $\frac{1}{2}$ & $\frac{1}{2(W_e+1)}$ & $\frac{\left(B_{e} + 1\right) \left(N + W_{e} + 1\right)}{2(B_{e}+1) \left( W_{e} + 1\right) + N \left(B_{e}+ W_{e} + 2 B_{e} + 2 W_{e} + 2\right) }$ \\
\rule{0pt}{6ex} $p_i$ & $\frac{1}{2}$ & $0$ & $\frac{\left(B_{e} + 2\right) \left(W_{e} + 1\right)}{2(B_{e}+1) \left( W_{e} + 1\right) + N \left(B_{e} W_{e} + 2 B_{e} + 2 W_{e} + 2\right) }$ \\
\bottomrule
\end{tabular}
\end{table*}

To quantify possible deviations from the competitive benchmark, we introduce the \emph{Price of Collusion} (PoC), which is analogous to the Price of Anarchy used in algorithmic game theory.

\begin{definition}[Price of Collusion]
We define the \textit{Price of Collusion} (PoC) as the ratio of social welfare under the competitive Nash equilibrium to that under the collusive equilibrium:
\begin{equation}
\text{PoC} = \frac{SW^{NE}}{SW^{Collusive}}.
\end{equation}
A value of $\text{PoC} > 1$ quantifies the proportional welfare degradation imposed on consumers due to multi-agent algorithmic collusion.
\end{definition}
The lower bound of PoC is between $0$ and $1$ under Bertrand congestion model with only entrants, depending on $W_e$ value.

\subsection{Revenue Expectations Under Learning}

In the preceding subsections, we summarized the complete-information Nash and collusive benchmarks for static congestion markets. Under multi-agent reinforcement learning, however, service providers repeatedly update their strategies, causing both prices and quantities to become stochastic processes. Consequently, a provider's revenue is also a random variable, and its expected value cannot generally be written as the product of the expected price and expected quantity because of the correlation induced by the learning dynamics.

The following results establish sufficient conditions under which this decomposition is valid for both Cournot and Bertrand competition. These identities provide a useful theoretical foundation for interpreting the empirical results in Section~IV, where market outcomes are obtained by averaging over repeated learning trajectories.

\begin{theorem}[Revenue Identity for Dynamic Cournot Competition]
\label{thm:generalized_revenue_identity}
If all active agents, including a market incumbent $i=1$ and competing entrants $2 \le i \le N+1$, maintain strictly non-zero prices and quantity allocations, the expected revenue $\mathbb{E}[r_i]$ for each firm satisfies the linear expectation identity:
\begin{equation}
\mathbb{E}[r_i] = \mathbb{E}[p_i]\mathbb{E}[x_i], \quad \forall i \in \{1, \dots, N+1\}.
\end{equation}
Specifically, this linear identity is preserved under the following conditions:
\begin{enumerate}
    \item Incumbent Condition: The identity holds for the incumbent ($i=1$) if its market supply variance is zero (i.e., operating at a deterministic, fixed supply baseline such that $\mathbb{E}[x_1^2] = \mathbb{E}[x_1]^2$).
    \item Entrant Condition: The identity holds for the competing entrants ($i > 1$) if the incumbent's strategic supply choices are orthogonal to and statistically independent of the entrant profiles (such that $\mathbb{E}[x_1 x_i] = \mathbb{E}[x_1]\mathbb{E}[x_i]$) and individual entrant supply variance is zero.
\end{enumerate}
Conversely, if any strategic profile results in a zero price or zero quantity, the linear expectation identity breaks down as the Wardrop conditions in (\ref{eq:wardropeq}) no longer hold.
\end{theorem}

\begin{theorem}[Revenue Identity for Dynamic Bertrand Competition]\label{thm:revenue_identity}
In the Bertrand congestion game with total entrants $\mathcal{N}$, if all competing entrants maintain the minimum market price $p_{\min}$ such that the set of active entrants satisfies $\mathcal{M}_{\min} = \mathcal{N}$, the expected revenue realized by each entrant $i \in \mathcal{M}_{\min}$ is identically equal to the product of its expected price and expected quantity allocation, given by:
\begin{equation}
\mathbb{E}[r_i] = \mathbb{E}[p_i]\mathbb{E}[x_i].
\end{equation}
Conversely, if an asymmetric market profile emerges where a subset of entrants strategically deviates by announcing nominal prices strictly exceeding $p_{\min}$, this linear expectation identity breaks down due to some providers serving no customers.
\end{theorem}

\section{Multi-Agent Reinforcement Learning}

The congestion games developed in Section II are repeated over multiple rounds and modeled as a Markov game in which each service provider acts as an independent reinforcement-learning agent. At each round, all providers simultaneously select either a price (Bertrand) or a quantity (Cournot), observe the resulting market outcome, receive their realized revenue as a reward, and update their policies. The agents do not know the underlying demand function and must instead learn effective strategies solely through repeated interaction with the environment.

Each agent's state consists of the joint actions taken during the previous $K$ rounds, providing a finite history of market behavior. The action space consists of a discrete set of allowable prices or quantities, depending on the competition model. The immediate reward for each agent is its realized market revenue,
\[
r_i = p_i x_i,
\]
where the allocated demand $x_i$ is determined by the Wardrop equilibrium described in Section II.

We primarily employ independent Deep Q-Network (DQN) agents to learn equilibrium strategies. Each agent maintains its own value network and replay buffer, while treating the remaining agents as part of the environment. Standard experience replay and target networks are used to stabilize training. To evaluate the robustness of the observed market behavior, we additionally compare DQN with an actor--critic method in the experimental section.

Algorithm~1 summarizes the overall learning procedure. During each episode, all agents simultaneously select actions according to an $\epsilon$-greedy exploration policy, observe the resulting revenues after market clearing, store the transition in replay memory, and update their neural-network parameters using mini-batch stochastic gradient descent. Training continues until the learned market behavior converges.

\subsection{Neural Network Architecture}

Each agent is represented by an independent multilayer perceptron with two hidden layers of 64 neurons. The input dimension depends on the history length $K$, while the output dimension equals the number of discrete actions. Unless otherwise stated, all experiments use the same network architecture and hyperparameters across both markets.

\subsection{Learning in Market with congestion}
The algorithm \ref{alg:bertrand_dqn} gives detailed market mechanism and training process for multi-agent deep reinforcement learning in market with congestion with DQN. 
\begin{breakablealgorithm}
\caption{Multi-Agent DQN in the market with Congestion}
\label{alg:bertrand_dqn}
\footnotesize
\begin{algorithmic}[1]
\REQUIRE State Dimension $K$, Action Set $\mathcal{A}$, Replay $\mathcal{D}_i$, Batch $B$, Discount Factor $\gamma$, Learning Rate $\alpha$.
\REQUIRE Congestible Resources: $B_e$ for Incumbent and $W_e$ for $N$ Entrants.
\STATE \textbf{Initialize:} Network $Q_{\theta_i}, Q_{\phi_i}$ ($\phi_i \leftarrow \theta_i$), $\mathcal{D}_i \leftarrow \emptyset$, State $\mathcal{S} = \langle \mathbf{a}^{(1)}, \dots, \mathbf{a}^{(K)} \rangle$
\FOR{Episode $e$ from $1$ \TO $E$}
    \STATE Select $a_i^e \in \mathcal{A}$ via $\epsilon$-greedy based on state $\mathbf{s} \leftarrow \bigoplus_{k=1}^K \mathbf{p}^{(k)}$
    \STATE \textbf{Market Execution:}
    \STATE Compute realized revenues based on action as current reward:
    \STATE \textbf{Learning and Updates:}
    \STATE Update state $\mathcal{S} \leftarrow \mathcal{S} \cup \{\mathbf{a}^{(e)}\} \setminus \{\mathbf{a}^{(1)}\}$ to construct next state $\mathbf{s}'$
    \FOR{each agent $i \in \{1, \dots, N+1\}$}
        \STATE Push transition $(\mathbf{s}, a_i, r_i, \mathbf{s}')$ into $\mathcal{D}_i$
        \IF{$|\mathcal{D}_i| \ge B$}
            \STATE Sample mini-batch $\mathcal{B}_i \sim \mathcal{D}_i$
            \STATE Compute TD target value: $y = r_i + \gamma \max_{a'} Q_{\phi_i}(\mathbf{s}', a')$
            \STATE Update $\theta_i$ via gradient descent:
            \STATE \quad $\theta_i \leftarrow \theta_i - \alpha \nabla_{\theta_i} \mathcal{L}(\theta_i)$
        \ENDIF
        \STATE Every $T$ episodes, update target network: $\phi_i \leftarrow \theta_i$
    \ENDFOR
    \STATE $\mathbf{s} \leftarrow \mathbf{s}'$, \quad $\epsilon \leftarrow \max(\epsilon_{\min}, \epsilon \cdot \epsilon_{\text{decay}})$
\ENDFOR
\end{algorithmic}
\end{breakablealgorithm}

\section{Empirical Results}
\subsection{Benchmark Validation}

\begin{figure}[htbp]
\centering
\includegraphics[width=\linewidth]{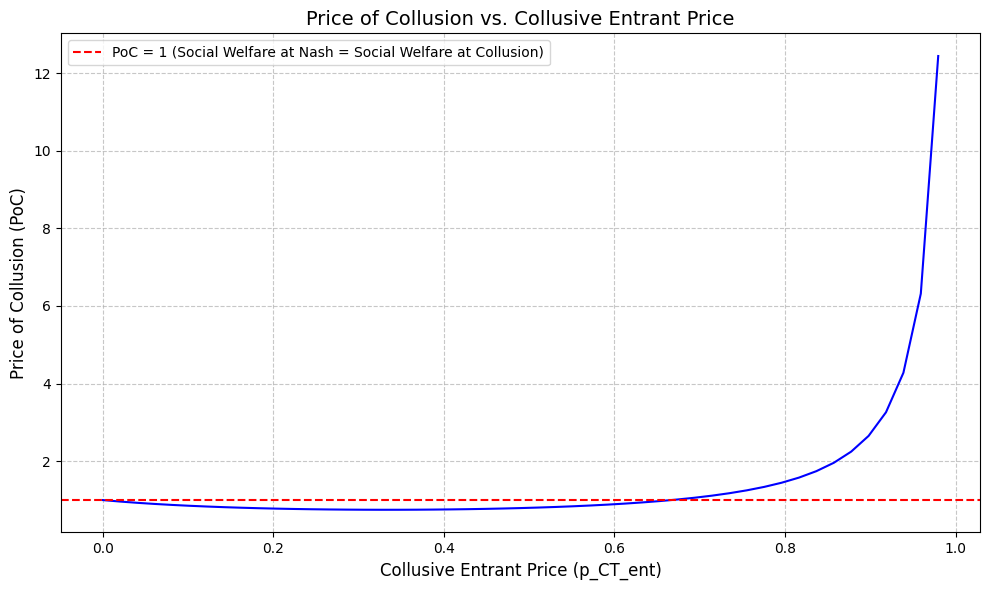}
\caption{Impact of colluding price on Price of Collusion (PoC) for Bertrand Congestion Model with only Entrants for $W_e=1$.}
\label{fig:poc}
\end{figure}

We first validate the analytical benchmarks introduced in Section III. Figure~\ref{fig:poc} illustrates the Price of Collusion (PoC) for the Bertrand congestion model without an incumbent ($W_e=1$) as the collusive price varies. The results demonstrate that tacit collusion does not necessarily reduce social welfare. When providers coordinate on relatively low prices, the resulting market allocation remains close to the competitive equilibrium while congestion is reduced, leading to a lower PoC. In contrast, collusion at higher prices substantially reduces consumer participation and increases the efficiency loss. As predicted by the analytical benchmark, the number of entrant providers has essentially no effect on the PoC.

These results demonstrate that the proposed PoC metric captures not only the existence of collusion but also its impact on overall market efficiency, providing a useful benchmark for interpreting the learning results that follow.

\subsection{Impact of State Representation}

We next investigate how the amount of historical information available to each learning agent influences the learned market behavior. Throughout this subsection we set $B_e=2$ and $W_e=1$. Each agent has ten discrete actions corresponding to uniformly spaced prices or quantities. Unless otherwise stated, all reported results are averaged over six independent Monte Carlo simulations.

\begin{figure*}[htbp]
\centering
\includegraphics[width=1.0\linewidth]{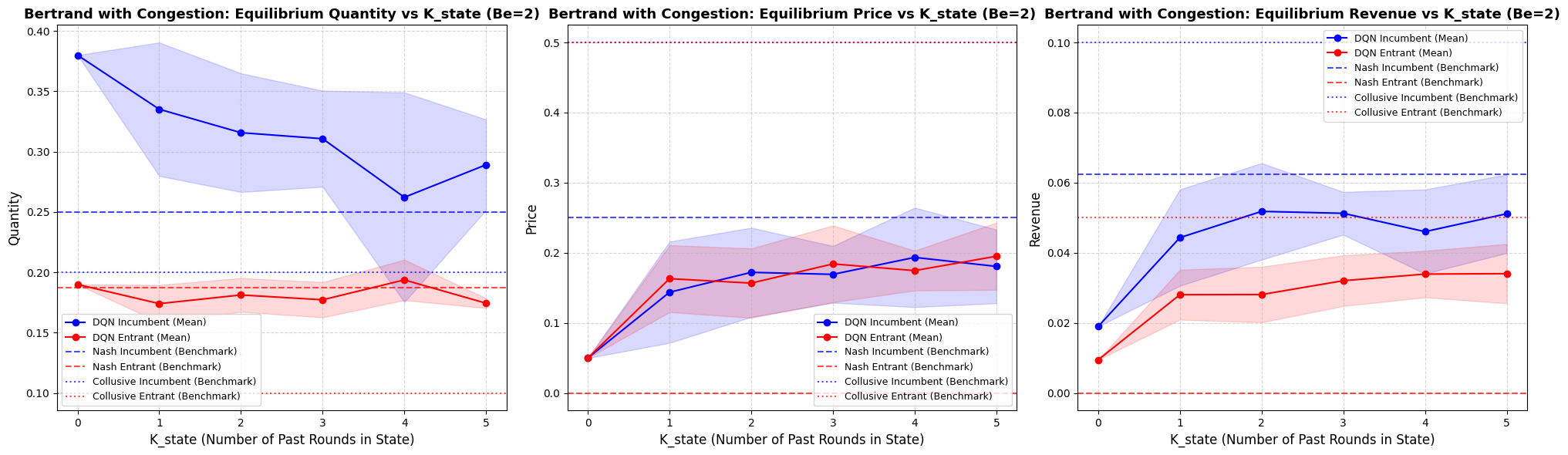}
\caption{Impact of State Design on the Learning Dynamics for Bertrand Congestion Model when $B_e=2$, $W_e=1$, $N=2$.}
\label{fig:state}
\end{figure*}

Figure~\ref{fig:state} shows the learned Bertrand prices for different history lengths $K$. When $K=0$, the state contains no historical information and the learning problem effectively reduces to a contextual bandit. In this case, the agents converge to the competitive Nash equilibrium by selecting the lowest feasible price. As the history length increases, however, the agents consistently learn higher prices corresponding to tacit collusion.

These results indicate that memory plays a critical role in enabling coordinated pricing behavior. Access to previous market actions allows the agents to condition future decisions on past interactions, thereby facilitating tacit collusion without explicit communication.

\subsection{Impact of Market Structure}

\begin{figure*}[htbp]
\centering
\includegraphics[width=\linewidth]{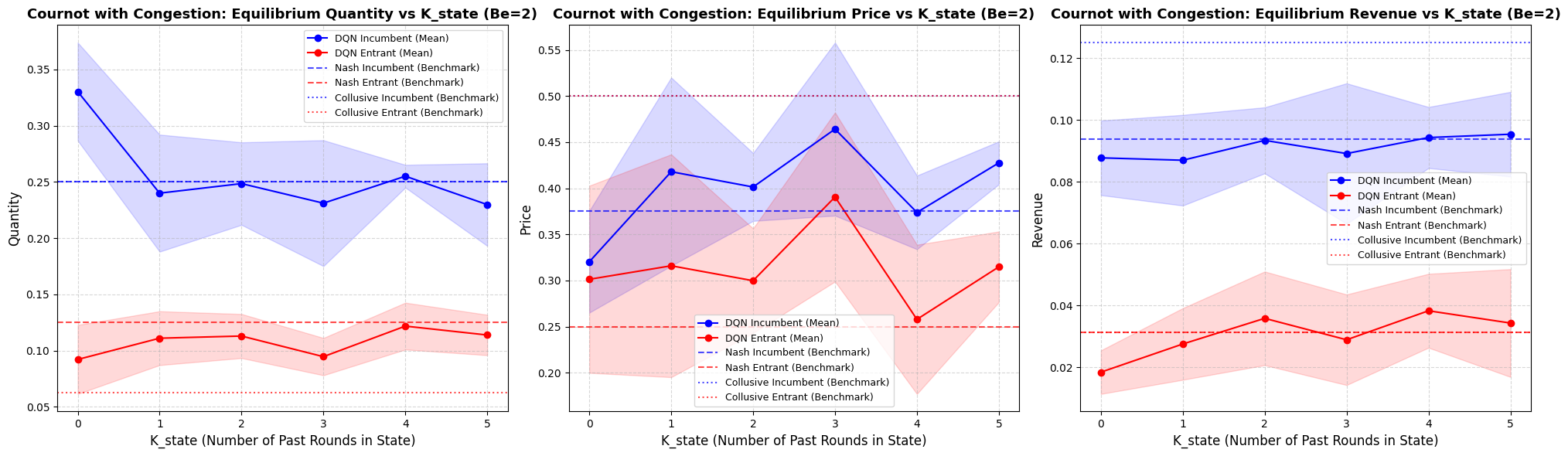}
\caption{Impact of Market Structure on the Learning Dynamics for Cournot Congestion Model when $B_e=2$, $W_e=1$, $N=2$.}
\label{fig:market}
\end{figure*}

Figure~\ref{fig:market} compares the learning dynamics under Bertrand and Cournot competition. The incumbent consistently converges to behavior close to the theoretical Nash equilibrium in both market models. The entrant providers, however, exhibit fundamentally different behavior.

Under Cournot competition, the learned strategies remain close to the theoretical equilibrium, consistent with the potential-game structure of the congestion model. In contrast, Bertrand competition produces substantially more collusive pricing behavior. Because Bertrand competition does not possess the same convergence guarantees as Cournot competition, repeated learning allows agents to sustain prices above the competitive equilibrium.

These observations demonstrate that congestion interacts strongly with the underlying market structure, leading to qualitatively different learning dynamics under price and quantity competition.

\subsection{Impact of the Incumbent}

We next examine the role of the incumbent by removing the licensed provider from the market and allowing only entrant providers to compete over the shared spectrum. We focus on the case of Bertrand competition.

\begin{figure*}[htbp]
\centering
\includegraphics[width=\linewidth]{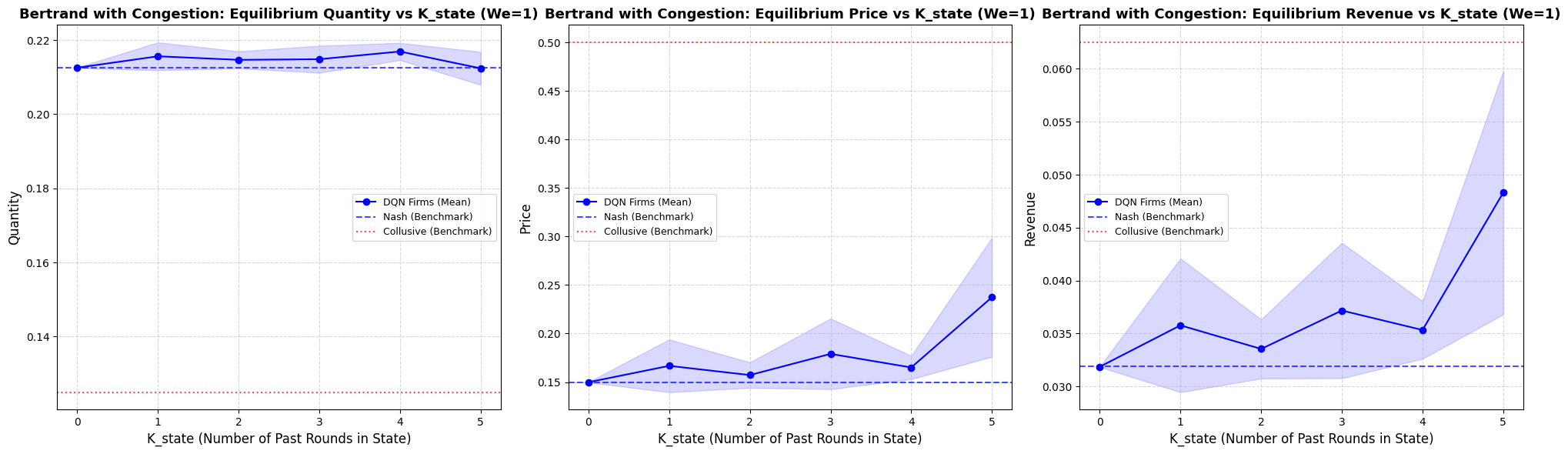}
\caption{Impact of Incumbent Removal on the Learning Dynamics for Bertrand Congestion Model with the First and Second Revenue Calculation Method when $W_e=1$, $N=2$.}
\label{fig:incumbent}
\end{figure*}

Figure~\ref{fig:incumbent} compares two methods for computing the average revenue of the entrant providers. The first computes the empirical average of realized revenues directly, while the second computes the product of the empirical average price and average quantity. As predicted by Theorem~2, the two approaches produce identical results under the conditions established by the theorem.

\begin{figure*}[htbp]
\centering
\includegraphics[width=\linewidth]{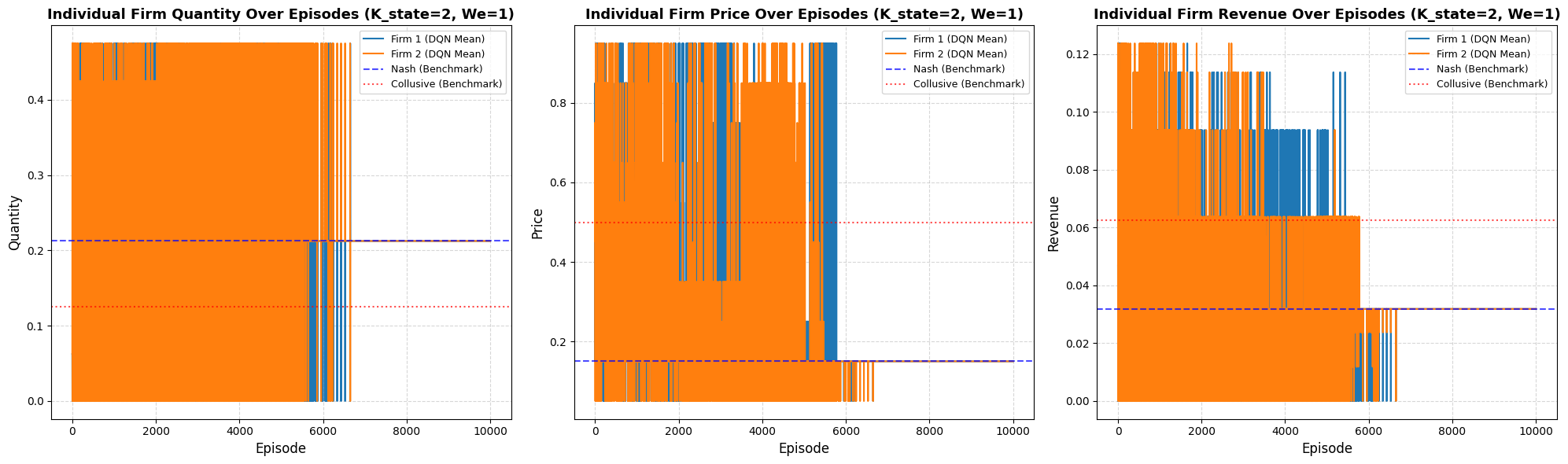}
\caption{Impact of Incumbent Removal on the Learning Dynamics for Bertrand Congestion Model over episodes when $W_e=1$, $N=2$, $K=2$.}
\label{fig:episodes}
\end{figure*}

Figure~\ref{fig:episodes} further illustrates the learning dynamics after removing the incumbent. In this setting, the learned equilibrium produces a PoC of approximately 0.886, indicating that the resulting tacit collusion slightly improves social welfare relative to the competitive benchmark. This counterintuitive behavior arises because the learned collusive prices remain relatively low while reducing congestion.

These experiments validate the theoretical revenue identities and illustrate how market composition influences the emergence of collusive behavior.

\subsection{Impact of Congestion}
Finally, we investigate how congestion influences the learned market outcomes. Tables~\ref{tab:bertrand} and~\ref{tab:cournot} compare the learned equilibria with 
the theoretical Nash and collusive benchmarks in the absence of congestion. The learned Cournot strategies remain close to the Nash equilibrium, whereas the Bertrand strategies lie between the Nash and collusive benchmarks, indicating partial tacit collusion.

Overall, these results demonstrate that congestion fundamentally alters the strategic incentives of learning agents. Depending on the underlying competition model, congestion can either reinforce competitive behavior or facilitate tacit collusion, highlighting the importance of incorporating congestion effects when analyzing autonomous wireless markets.

\begin{table}[h]
\centering
\footnotesize
\caption{Empirical Results for Classic Bertrand Model.}
\label{tab:bertrand}
\setlength{\tabcolsep}{0.02\linewidth} % Restores clean, balanced horizontal spacing
\begin{tabular}{|c| c| c| c|}
\toprule
{\tt Metric} & {\tt Learned: Mean and Std} & {\tt Nash} & {\tt Collusive} \\
\midrule
$x_1$ & 0.416 (0.063) & \textbf{0.500} & 0.250 \\
$x_n$ & 0.424 (0.069) & \textbf{0.500} & 0.250 \\
$p_1$ & 0.233 (0.074) & \textbf{0.000} & 0.500 \\
$p_n$ & 0.249 (0.123) & \textbf{0.000} & 0.500 \\
$r_1$ & 0.067 (0.015) & 0.000 & \textbf{0.125} \\
$r_n$ & 0.063 (0.013) & 0.000 & \textbf{0.125} \\
\bottomrule
\end{tabular}
\end{table}

\begin{table}[h]
\centering
\footnotesize
\caption{Empirical Results for Classic Cournot Model.}
\label{tab:cournot}
\setlength{\tabcolsep}{0.02\linewidth} % Balanced horizontal padding for a clean fit
\begin{tabular}{|c| c| c| c|}
\toprule
{\tt Metric} & {\tt Learned: Mean and Std} & {\tt Nash} & {\tt Collusive} \\
\midrule
$x_1$ & 0.335 (0.040) & \textbf{0.333} & 0.250 \\
$x_n$ & 0.331 (0.069) & \textbf{0.333} & 0.250 \\
$p_1$ & 0.334 (0.056) & \textbf{0.333} & 0.500 \\
$p_n$ & 0.334 (0.056) & \textbf{0.333} & 0.500 \\
$r_1$ & 0.108 (0.025) & \textbf{0.111} & 0.125 \\
$r_n$ & 0.103 (0.016) & \textbf{0.111} & 0.125 \\
\bottomrule
\end{tabular}
\end{table}

\section{Conclusion}
This paper investigated how autonomous service providers learn competitive strategies in congestion-based spectrum markets when market demand is unknown. We introduced a multi-agent reinforcement learning (MARL) framework for this setting. By removing the classical assumption of complete demand knowledge, we evaluated how adaptive learning algorithms navigate congestion externalities across both Bertrand and Cournot competition. Our findings demonstrate a fundamental divide: while Cournot competition reliably converges toward theoretical Nash equilibria due to its potential game properties, Bertrand competition bypasses the classical zero-price paradox and fosters tacit algorithmic collusion. Future directions include exploring regulatory mechanisms to mitigate algorithmic collusion.

\clearpage
\bibliographystyle{plain}
\bibliography{ref}
%%%%%%%%%%%%%%%%%%%%%%%%%%%%%%%%%%%%%%%%%%%%%%%%%%%%%%%%%%%%
\clearpage
\appendix

\end{document}